\documentstyle[multicol,aps]{revtex}
\begin{document}
\draft
\input epsf

\title{Power Laws in Solar Flares: Self-Organized Criticality or 
Turbulence?}

\author{Guido Boffetta${}^1$, Vincenzo Carbone${}^2$, Paolo Giuliani${}^2$,
Pierluigi Veltri${}^2$
and Angelo Vulpiani${}^3$}
\address{${}^1$Dipartimento di Fisica Generale
 and  Istituto Nazionale di Fisica della Materia, 
Universit\'a di Torino \\
 Via Pietro Giuria 1, 10125 Torino, \\
Istituto di Cosmogeofisica - CNR, Corso Fiume 4, 10133 Torino, Italy}
\address{${}^2$ Dipartimento di Fisica
 and  Istituto Nazionale di Fisica della Materia, 
 Universit\'a della Calabria, 
\\  87036 Roges di Rende, Italy}
\address{${}^3$ Dipartimento di Fisica,
 and  Istituto Nazionale di Fisica della Materia, 
 Universit\'a "La Sapienza" \\
 Piazzale A. Moro 2, 00185 Roma, Italy}
\date{\today}
\maketitle

\begin{abstract}

We study the time evolution of Solar Flares activity by looking
at the statistics of quiescent times $\tau_{L}$ between 
successive bursts. 
The analysis of 20 years of data reveals a power law distribution
with exponent $\alpha \simeq 2.4$
which is an indication of complex dynamics with long correlation times.
The observed scaling behavior is in contradiction with the 
Self-Organized Criticality models of Solar Flares which predict
Poisson-like statistics.
Chaotic models, including the destabilization of the laminar phases
and subsequent restabilization due to nonlinear dynamics, are able to
reproduce the power law for the quiescent times. In the case of
the more realistic Shell Model of MHD turbulence we are able to reproduce 
all the observed distributions.
\end{abstract}


\begin{multicols}{2}

Solar flares are sudden, transient energy release above active 
regions of the sun \cite{Priest}. Energy is released 
in various form (thermal soft X--ray emission, accelerated particles, hard 
X--ray (HXR) emission, and so on). Parker \cite{Parker}
conjectured that 
flares represent the dissipation at
the many tangential discontinuities arising 
spontaneously in the bipolar fields of the active regions of the Sun 
as a consequence of random continuous motion of the footpoints of the field in 
the photospheric convection \cite{Parker}. Probability distributions,
calculated for various observed quantities, $x$, 
can be well represented by power 
laws of the form $P(x) = A x^{-\alpha}$.
In particular from HXR emission, the distribution of 
peak flux yields $\alpha \simeq 1.7$, that of total energy associated with a 
single event yields $\alpha \simeq 1.5$ and finally the distribution of 
flare duration yields $\alpha \simeq 2$ \cite{esperimenti}. 

The conjecture by Parker and the power laws found in the distributions 
of real flares stimulated a new way of looking at impulsive events like 
flares. In fact Lu and Hamilton \cite{Luetal} pointed out that 
Self--Organized Criticality (SOC), introduced earlier \cite{panzone}, could 
describe the main features of HXR flares, \cite{SOC,SOC2}, even if recent 2D MHD simulations have been 
devoted to recover power laws in the energy dissipation \cite{simula}. What is 
usually called SOC is a mechanism of charging and discharging, apparently 
without tuning parameters, which reproduces self--similarity in critical 
phenomena.

Lu and Hamilton, \cite{Luetal}, assume that the coronal magnetic field
evolves 
in a self--organized critical state in which an "event" can give rise to 
other similar "events" through an avalanche process (sandpile model). 
An active region on the sun is thus modeled through a 
magnetic field ${\bbox B}$ on a uniform 3D lattice. 
In order to have a statistically stationary state, energy is injected into 
the system by adding a small magnetic field increment $\delta {\bbox B}$
at a random site on the grid.
When an avalanche takes place, the energy input is suspended until
all sites become stable. In this sense the avalanches (flares) are 
fast phenomena, on a time scale much smaller than the injection 
mechanism. The large 
popularity of these models settles on their capability to reproduce
the power law behavior in the distribution functions of the total
energy, the peak luminosity and the duration of avalanches.

\begin{figure}[ht]
\epsfxsize=220pt\epsfysize=175pt\epsfbox{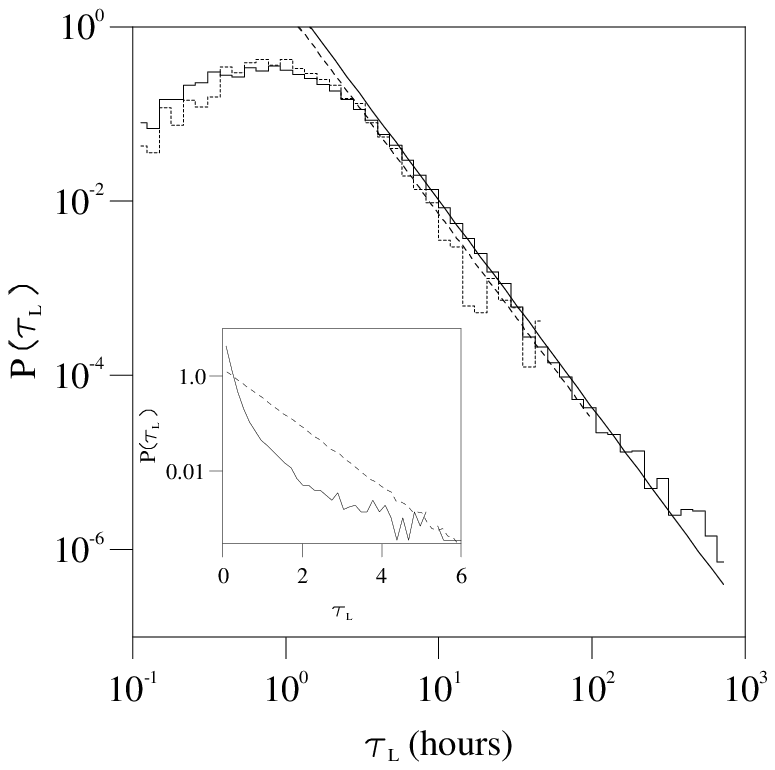}
\caption{
Probability distribution of the laminar time $P(\tau_L)$ between two X--ray
flares for dataset A (dashed line) and dataset B (full line). The straight
lines are the respective power law fits.
In the inset we show, in lin--log scale, the distribution for dataset B  
(full line) and the distribution obtained through the SOC model 
(dashed line) which displays a clear exponential law. The 
variables shown in the inset have been normalized to the respective 
root--mean--square values.
}
\label{fig1}
\end{figure}

What we want to stress in this letter is the fact that a different kind
of statistics can be studied on solar flare signals: the distribution of 
laminar or waiting times, i.e. the time intervals between two successive 
bursts. This distribution has been recently studied on solar flares  
HXR events \cite{laminari1,laminari2}. Wheatland et al. 
\cite{laminari2} have also emphasized the fact that this kind of 
distribution is crucial from the point of view of the avalanche model. SOC 
models indeed are expected to display an exponential waiting time distribution 
$P(\tau_{L}) = \langle\tau_L\rangle^{-1} \exp(-\tau_L/\langle\tau_L\rangle)$, 
where $\langle\tau_L\rangle$, the 
average laminar time, depends on the parameters of the model. This behavior is 
related to the fact that the avalanche duration is much smaller 
than the charging time (the time between two successive throws of magnetic
field in random position) and charging place is independent on the avalanche
position. Then one expects no correlation between successive bursts and
thus a trivial statistics for the laminar times. This is clearly observed 
in a simulation of the SOC automaton that we have done 
(see the inset of figure \ref{fig1}). On the contrary all authors 
\cite{laminari1,laminari2} found a 
more or less well defined power law distribution. In particular Wheatland 
et al. \cite{laminari2}, by performing a careful statistical analysis of 
waiting time distribution on 8 years of solar flares HXR bursts observed by 
the ICE/ISEE 3 spacecraft, have shown that the distribution in no way can 
be attributed to a nonstationary Poisson process. 

We have done the same statistics on laminar phases using twenty years of 
data from National Geophysical Data Center of USA. In this database 
starting and ending times as well as peak times of HXR bursts associated 
to flares from 1976 up to 1996, measured at the Earth by satellites in the 
$0.1$ to $0.8$ nm band, are stored. We calculated the laminar times as the 
time differences between two successive maxima of the flares intensity 
recorded during periods of activity of the same instrument. 
Two different kind of analysis have been performed: in the first one we have 
built up a dataset of about $1100$ samples (hence on dataset A)
by calculating only the 
differences between the time of occurrence of flares within the same 
active region, as identified through the H$\alpha$ flares occurrence. To 
build up the second dataset (dataset B),
we have considered the sun as a unique physical 
system, and we have calculated the time differences between two successive 
maxima of flare intensity regardless of the position of the flare on the 
sun surface. In this way we get a 
dataset of about $32,000$ samples. The analysis of these data shows that, 
in both cases, laminar times display a clear power--law distribution 
$P(\tau_{L})=A\tau_{L}^{-\alpha}$ 
with $\alpha = 2.38 \pm 0.03 $ 
in the range $6$ hr $\leq \tau_L \leq 
67$ hr (reduced $\chi^2=2.2$) for dataset B 
and $\alpha = 2.4 \pm 0.1$ (reduced $\chi^2 =1.1$) for 
dataset A(figure \ref{fig1}). 
The exact value of the exponent can be affected by the finite lenght of the 
observation times,
which underestimates the occurrence of long waiting 
times. However these results, as well as those obtained by previous authors 
\cite{laminari1,laminari2}, allow us to 
consider the power law distribution of waiting times as firmly established as 
the power laws observed for total energy, peak luminosity and time duration 
and force us to investigate whether models different from SOC can account 
for all these distributions.

One point of the SOC philosophy is that the system has to be
at the edge of chaos in order to display power laws
\cite{Chenetal90}. 
Indeed in a chaotic system there exists a characteristic time $t_c$
given in terms of the leading Lyapunov exponent as
$t_c\simeq 1/\lambda$. This seems to be incompatible with
the existence of self-similar scaling laws (i.e. no characteristic times)
a part the limit case $\lambda=0$.
The above argument in not very strong. As a matter of fact we recall
that it is possible to have chaotic systems showing scaling behavior
in presence of many time scales \cite{Aurelletal96} or in 
presence of strong
fluctuations of the local Lyapunov exponent \cite{Crisantietal92}. Thus it is 
worth investigating which kind of chaotic systems can give rise to 
self-similar scaling laws.

It is worth noting that the occurrence of a power law in the distribution of 
the laminar times is the analogous for the Solar flares of the Omori's law 
for the earthquakes \cite{Omo} and represents a clear indication of the 
existence, in the flare dynamics of strong correlations between successive 
bursts, at variance with the SOC model. 
The unique possible origin of the correlations  arises from 
non trivial evolution equations of the phenomenon. Because solar flares 
are governed by MHD equations, it is rather natural to investigate their
statistics in the context of MHD turbulence. Turbulence is a common phenomenon
in fluids, where chaotic dynamics and power law statistics coexist. 
Moreover fluid turbulence displays time intermittency, in that dissipative 
events are not uniformly but burstly distributed in time. 

Before introducing the turbulence model, let us stress the fact that the 
intermittent behavior can be observed also in simple dynamical models. 
Let us consider the one dimensional random map $x_{t+1} = r_{t} 
x_{t} (1-x_{t})$, where at each step $t$ the random variable $r_t$ is 
extracted according
to a given distribution, e.g. $r_t=4$ with probability $p$ and $r_t=1/2$
with probability $1-p$. If $p<1/3$
$x=0$ is an attracting fixed point.
If $p=1/3+\delta p$ one has a rather interesting behavior called
on-off intermittency \cite{PST93}: $x_t$ remains close to zero for a
certain time (laminar time) then there is a short interval of strong 
activity (burst) after which a new quiescent phase takes place. 
Also in this simple model, if $\delta p$ is not too large, one observes 
power law not only for the energy distribution (here defined as the integral
of $x_t$ over the burst), but also for the laminar times $\tau_L$. 
The exponent of power law for $\tau_L$ turns out to be $\alpha = 1.5$ 
\cite{onoff}. 
In spite of its simplicity, the random map model contains some basic
ingredients of the relevant features of the time intermittency in 
dynamical systems, i.e. the 
destabilization of the laminar phase by linear instability and the 
subsequent restabilization due to the nonlinear dynamics.

Dynamical systems more directly related to fluid turbulence are the 
so called shell models \cite{BJPV98}. Shell models represent 
a zero-order approximation of fluid equations (Navier-Stokes
or MHD equations) in which one consider a single (complex) scalar variable
$u_n$ (and $b_n$) as representative of the velocity (magnetic)
fluctuation associated to a wavenumber $k_n=k_0 2^n$ ($n=1, ...,  N$).
The fact that the wavenumbers $k_n$ are exponentially spaced allows
to reach very large Reynolds numbers with a moderate number of
degrees of freedom and then to investigate regimes of $3D$ MHD turbulence 
which are not accessible by direct numerical simulation. The philosophy 
underlying the shell model approach to turbulence 
is that even with a relatively small dynamical system it is 
possible to reproduce some statistical features of the turbulent
cascade. In particular shell models mimic at the best the time 
intermittency of real turbulent fluid flows.

The evolution equations for the dynamical variables $u_n$ and $b_n$ are 
built up by ignoring any detail of the spatial structure and boundary 
conditions. Only the interactions between nearest and next nearest 
neighbor shells are retained 
in the form of quadratic nonlinearities. The coupling 
coefficients of nonlinear terms are determined by imposing 
the inviscid conservation of the MHD quadratic invariants
\cite{BJPV98,JPV91}. The particular shell model we used in 
our simulation reads \cite{GC98}
\begin{eqnarray*}
{d u_n \over dt} = - \nu k_n^2 u_n +f_n+
ik_n \left\{(u_{n+1}u_{n+2} -b_{n+1} b_{n+2}\right) -\nonumber \\
{1 \over 4} \left(u_{n-1}u_{n+1} - b_{n-1} b_{n+1}\right)-
{1 \over 8} \left(u_{n-2}u_{n-1} - 
b_{n-2} b_{n-1}\right)\}^{*} 
\end{eqnarray*}
\begin{eqnarray*}
{d b_n \over dt} = - \eta k_n^2 b_n +
ik_n ({1/6})\left\{  \left(u_{n+1}b_{n+2} - 
b_{n+1} u_{n+2}\right) + \right.  \nonumber \\
\left(u_{n-1}b_{n+1} - b_{n-1} u_{n+1}\right)
+ \left.  \left(u_{n-2}b_{n-1} - 
b_{n-2}u_{n-1}\right)\right\}^{*}
\end{eqnarray*}
where $\nu$ and $\eta$ are respectively the viscosity and the resistivity
and $f_n$ is an external forcing term acting only on velocity fluctuations.

Shell models are good models of turbulent cascade in the sense that they 
display, in the limit of fully developed turbulence $\nu, \eta \to 0$, scaling 
laws for the structure functions $S_p(n) = \langle |x_n|^p \rangle \sim 
k_n^{-\zeta_p}$ where $x_n$ is either $u_n$ or $b_n$. In the hydrodynamic 
limit ($b_n=0$) the set of scaling 
exponents $\zeta_p$ are found to be very close to those obtained 
by experiments \cite{JPV91}. One indeed observes a clear deviation
from the Kolmogorov scaling ($\zeta_p=p/3$) as a consequence
of the intermittent dynamics of the system.

\begin{figure}[ht]
\epsfxsize=220pt\epsfysize=175pt\epsfbox{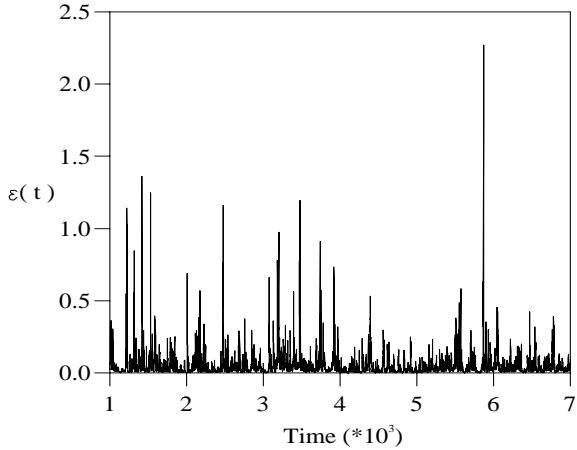}
\caption{
Time series of energy dissipation $\epsilon(t)$ for the shell
model. The parameters used in the simulation are
$N = 19$, $\nu = \eta = 10^{-7}$, $k_0 = 1$. The external forcing term $f_n$ 
is a stochastic variable acting only on the first two shells of the 
velocity fluctuations. 
It is calculated according to the Langevin equation 
$df_n/dt = -f_n/\tau_0 + \mu$, where $\tau_0$ is the characteristic time of 
the largest shells ($\tau_0 \simeq 10$ in our units) and $\mu$ is a 
Gaussian white--noise with $\sigma = 0.1$. 
}
\label{fig2}
\end{figure}

Another observable whose statistics is well reproduced in shell
models is the energy dissipation $\epsilon(t)$ 
defined as 
$ \epsilon(t) = \nu \sum_{n=1}^{N} k_n^2 |u_n|^2 +
\eta \sum_{n=1}^{N} k_n^2 |b_n|^2
$
which displays the characteristic intermittency of fully developed
turbulence (see figure \ref{fig2}).

\begin{figure}[ht]
\epsfxsize=220pt\epsfysize=175pt\epsfbox{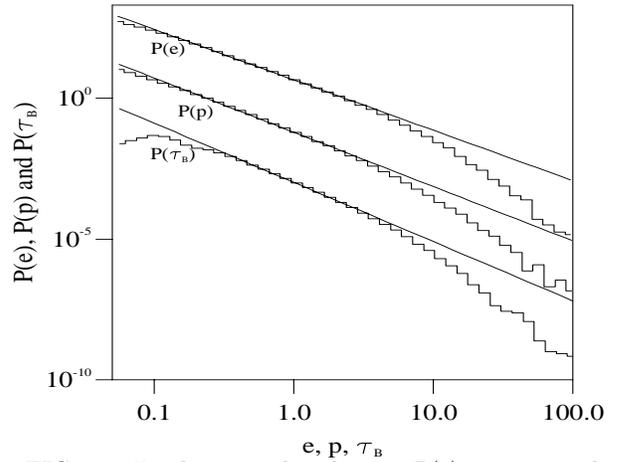}
\caption{
Total energy distribution $P(e)$, energy peak distribution $P(p)$ and 
bursts duration distribution $P(\tau_B)$ for the shell model. The variables 
have been normalized to the respective root--mean--square values. 
The straight lines are the fits with power laws.
The values of $P(e)$ and $P(\tau_B)$ are offset by a factor $100$ and $10^{-2}$ 
respectively.
}
\label{fig3}
\end{figure}

\begin{figure}[ht]
\epsfxsize=220pt\epsfysize=175pt\epsfbox{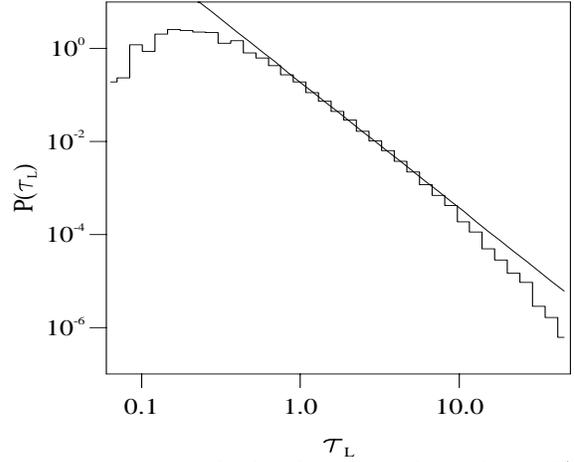}
\caption{
The distribution of laminar times $P(\tau_L)$ for the Shell model, normalized 
to the root--mean--square. The straight line is the fit with a power law.
}
\label{fig4}
\end{figure}

We have performed on $\epsilon(t)$
the same statistical analysis done for the solar flare signal. 
We define a burst of dissipation (corresponding to a flare) 
by the condition $\epsilon(t)\ge \epsilon_c$. This definition allows us to 
calculate the distribution functions for the peak values of the bursts, their 
total energy (defined 
as the integral of the signal above $\epsilon_c$) and the duration of the 
bursts (defined as the time during which the dissipation is above
$\epsilon_c$). We have chosen the threshold
as $\epsilon_c = \langle\,\epsilon(t)\,\rangle + 2 \sigma$, where the
average and the standard deviation have been calculated 
on the time intervals in between the bursts, through an iterative
process in order to take
into account only the background contribution. The results of the analysis
are shown in figure \ref{fig3}. 
Also in this case we observe clear power law 
distribution functions with exponents $\alpha \simeq 2.05 $ for 
the peak distribution, $\alpha \simeq 1.8$ for the total energy distribution 
and $\alpha \simeq 2.2$ for the burst durations. The exponents are close
to those obtained in analyzing solar flares data, but we do not think that the 
agreement is particularly significant as the model exponents depend
on the value chosen for the threshold. The relevant point is that, at variance 
with SOC models, MHD shell models display a power law statistics also for the 
laminar times, as shown in figure \ref{fig4}. The scaling exponent turns out 
to be $\alpha \simeq 2.70$, close to the one 
obtained from the experimental data. 

The different behavior of SOC models and turbulent MHD shell models is related 
to the conceptually different mechanisms underlying the SOC phenomenon 
and the phenomenon of intermittency in fully developed turbulence. SOC models
represent self-similar phenomena, while the intermittent behavior of turbulence 
is related to its chaotic nature \cite{JPV91}. Let us stress again that the 
actual value of the numerical scaling exponent is not important
as it could depend on the details of the model. The statistics of the
laminar times between two bursts is due to global properties 
of the system and it is thus more relevant than the properties
of the single burst. In this sense, a good model for the flare bursts 
should be able to reproduce the quiescent time distributions. SOC models 
predict a Poissonian statistics for the scaling behavior of the quiescent times
in the Solar flares activity. On the contrary chaotic models
are able to reproduce the power law for the quiescent time since they  
include the correct mechanism for destabilization of the laminar phases
and subsequent nonlinear restabilization.

We thank G. Caldarelli and A. Vespignani for useful discussions.
This work has been partially supported by INFM (PRA-Turbo), 
 MURST (no. 9702265437),  the European Network 
{\it Intermittency in Turbulent Systems} (contract number FMRX-CT98-0175),
CNR (Special Project {\it Turbulence and nonlinear phenomena in plasmas})
and contract no. 98.00148.CT02, and Agenzia Spaziale Italiana (ASI) contract
no. ARS 98-82.


\newpage
\narrowtext

\end{multicols}
\end{document}